r4xian@uwaterloo.ca; MSc Candidate                                        Xian   1

# A Comparative Study of Disordered and Ordered Protein Folding Dynamics Using Computational Simulation

**Abstract**

Folding protein dynamics has been an area of high interest for quite some time, especially given the increased focus on the field of Biophysics. Because folding dynamics occur on such short time scales, empirical techniques developed for more "static" protein events, such as x-ray crystallography, nuclear magnetic resonance, and GFP labelling, aren't as applicable. Instead, computational methods must often be used to simulate these short lived yet highly dynamic events. One such computational method that's proven to provide much valuable insight into protein folding dynamics is Molecular Dynamics Simulation (MD Simulation). This simulation method is both highly computationally demanding, yet highly accurate in its modelling of a protein's physical behaviour. Besides MD Simulation, simulations in general, are quite applicable in the context of these protein events. For example, the simple Gillespie algorithm, a computational technique which can be executed on almost any personal computer, provides quite the robust view into protein dynamics given its computational simplicity. This paper will compare the results of two simulations, an MD simulation of a disordered six residue fragment, and a Gillespie algorithm based simulation of an ordered folding protein, the mathematically identical nature of the Gillespie algorithm results of the asymptotically stochastic arc tangent dynamics for the wild type predicting the exact behaviour of the carcinogenic protein system concentration dynamics shows the computational power simulations provide for analyzing both disordered and ordered protein systems.

**Introduction**

Proteins are one of three major classes of biological macromolecules. Connected by peptide chains, these long chains of amino acids are essential for cellular signalling, ligand binding, and general bodily functioning. Broadly, there exists two classes of proteins: ordered and disordered. The names in their simplicity are telling. Ordered proteins are those proteins with discrete, well-defined 3D (or tertiary) structures. Their conformations in space generally change with a clear trigger—a binding event, a pH change, a temperature change. However, disordered proteins possess no such structural order. In fact, their 3D conformation constantly interconverts between a series of energetically favourable macro states. Even more intriguingly, these interconversions seem to follow no discernible pattern, nor are they triggered by any specific event; they interconvert stochastically and spontaneously. Their disorder is so embedded into their behaviour that it can be seen as intrinsic, coining the term intrinsically disordered proteins (IDPs).

IDPs are so unconventional in their behaviour that they defy one of the most ubiquitous protein paradigms: the protein folding paradigm. It's been long thought that the primary sequence, the sequential order of the amino acids in a protein, fully determines its structure and thus its function. Though this paradigm holds true for ordered proteins, its veracity fails when applied to IDPs. For though it's true that generally, structure determine function, how can one possibly determine the function of an IDP if its structure is never constant? Consequently, Scientists often apply more mesoscopic techniques to describe IDPS: RMSD, radius of gyration, number of hydrogen bonds. Beyond the difference in the experimental and analytic techniques applied to disordered and ordered proteins, there exists a clear difference in the computational methods applied to both. The basic methodology of MD simulation is the application of a potential energy function, force field, to inform the movement of the simulated system. This force field is generally a summation of many components, each carefully formulated to emulate some biophysically realistic feature. As the force field acts on the system throughout each time step, the system's coordinates of motion are updated accordingly and fed back into the force field, which then integrates Newton's 2nd law in order to reapply the process for the next time step. The relatively simpler computational methods used for ordered proteins, though more basic in their computational logic and power, can sometimes involve more computational understanding in order to execute them. The method used to model the ordered folding protein of this paper was the Gillespie Algorithm. Its methodology is based in chemical kinetics, implementing a combination of probabilistic dynamical formulas and stochastic time step determination. A detailed methodology of both computational methods will be carefully outlined in the next section.

**Methods**

The disordered MD Simulation carried out involved a six-residue fragment of the alpha-beta peptide KLVFFA, using the CHARM36m force field and CHARM-modified TIP3P water model. First, the amino acid sequence was entered into UCSF Chimera to manually generate the PDB file. The peptide was then centred in a rhombic dodecahedron simulation box with a 1.5 nm box edge separation, and again solvated with spc216.gro, then neutralized with 150mM NaCl. The single peptide was then energy minimized using the steepest descent algorithm, with a 0.95 nm cut-off for short-range electrostatic and Lennard Jones interactions. Following energy minimization, the system was submitted for a high temperature production run of 450 K under 500 ps. From the resulting production run, eight configurations from unique time steps were randomly chosen, then used to construct an eight-peptide, 2x2x2 system (see Figures for visualization); the conformational sampling of the peptide under high temperature emulated the rapid structural interconversion of IDPs, with each of the eight chosen configurations representing unique IDP conformations. The eight-peptide system was then again centred in a rhombic dodecahedron box with 0.1 nm box edge spacing and energy minimized using the same parameters as the minimization of the single-peptide system. Afterwards, NPT equilibration of the eight-peptide system was performed in two steps, first with Berendsen pressure coupling, followed by Parrinello-Rahman barostat application, both for 100 ps. With the completion of NPT equilibration, the eight-peptide system was finally ready for production run, in which a 990 ns long run was conducted.

As for the ordered protein simulation, a two-state system was chosen to be simulated with Python 3.7.1: an unidentified protein that can only exist in either a folded or unfolded configuration. The number of folded proteins was initially set to be zero to emulate a completely entropic system of unfolded proteins achieving some steady state concentration of folded proteins. The rate of conversion between unfolded to folded, k_fold, and vice versa, k_unfold (supporting Figure 1), were equally set to be 110 seconds^-1 to mimic a freely folding and unfolding protein unaided by rate-increasing enzymes. The volume of the box was set to be 100 length units^3, and the total number of proteins N in the system was set to 10,000. The total number of time steps T_steps was 100,000. A counter T was defined to keep track of real time after each simulation iteration. The last step before the algorithm could start iteration was the definition of two empty arrays T_array and N_array to store the real time T after each time step del_t and the number of folded molecules N_folded.

The simulation for loop was iterated over T_steps with the Gillespie Algorithm, which executed the following process each time it iterated through a time step: given the current number of folded proteins N_folded, the propensities for folding and unfolding were calculated as functions of k_fold, k_unfold, total number of proteins N, and N_folded using the Master Gillespie Equations for propensity (Caltech, n.d.). Then, a random probability prob_r was generated from np.random.random to simulate the stochastic process: if prob_r was less than the calculated folded propensity P_fold, the number of folded proteins N_folded decreased by one, otherwise, N_folded increased by one (Caltech, n.d.). Then, to choose the time step to update time T by, a random value from an exponential distribution np.random.exponential was chosen, in which the argument of the exponential tau is defined as tau=1.0/(k_fold_tilde + k_unfold_tilde), where k_fold_tilde and k_unfold_tilde have units of number of molecules/second and are defined by the product of number of molecules at that time step times the rate of conversion to that molecule (Caltech, n.d.). Finally, the real time T was then updated by the time step del_t generated from np.random.exponential(tau), and the algorithm looped again to execute the same process for each time step in T_steps, this time with a new protein ensemble of folded and unfolded proteins (Caltech, n.d.). After the simulation ran for its designated 100,000 time steps, it was seen that the data it produced corresponded to about 0.05 seconds in real time, a high pay-off for MD simulation.

After production run, the disordered eight peptide fragment was submitted for two classes of mesoscopic analysis: RMSD, radius of gyration, and number of Hydrogen Bonds. All these quantities are rough measures of the fluctuation in 3D structure and can be seen below. As for the ordered folding protein, the simulated concentration data of folded proteins N_array/V was then plotted against T_array, the real time, to view the trajectory of the folded protein concentration equilibration. This trajectory can also be seen below. With the increase in the concentration of k_fold while maintaining k_unfold, one can see the clear approach of the ordered system to the behaviour of the disordered RMSD plot.

**Discussion**

Qualitatively, the most striking feature of the comparison between the simulation results of the ordered and disordered system is the similarity in the behaviour of the RMSD graph for the eight peptide system, and the concentration versus time graph of the folded ordered proteins. This similarity is the most overarching conclusion that can be drawn from this study, and will be discussed in greater detail near the end of the discussion. First, the results of the Gillespie algorithm can be analyzed quite simply. From the mother ODE of chemical kinetics (which can be seen under the Relevant Equations section), the resulting equation for time-dependent concentration is a hyperbolic tangent of variables time t, rate of folding k_fold, rate of unfolding k_unfold, and volume v. Though the tanh function is sigmoidal, physically one can't have negative time; thus, we can only observe the positive time portion of the folding dynamics. As one increases the rate of folding k_fold while keeping k_unfold constant, one can see the tanh behaviour isn't lost, however the noise of the equilibrating portion increases dramatically. As well, the time taken to equilibrate decreases as k_fold increases. When k_fold and k_unfold are equal, the concentration equilibrates at around 0.02 seconds, approaching around 50 folded proteins at a steady state. When k_fold=10(k_unfold), the concentration seems to equilibrate at around 0.005 seconds, reaching a steady state of nine folded proteins. As for k_fold=100(k_unfold), the system equilibrates even faster, almost instantaneously, reaching a stead state of only one folded protein. Clearly in this case, the graph of c(t) resembles the RMSD graph virtually identically. As well, the relative amplitude of the equilibrating fluctuation increases as k_fold increases, with a relatively smooth equilibration portion when the folding and unfolding rates are equal, a noticeable fluctuation amplitude of around one folded protein when k_fold=10(k_unfold) (when the steady state value is about nine folded proteins), and an extremely high fluctuation amplitude of around 0.4 folded proteins (given the steady state folded protein value of around one protein).

The radius of gyration and number of hydrogen bond data for the disordered eight peptide system provide results consistent with what is most coarsely known for IDPs. Throughout the 10 ps run of the eight-peptide system, the radius of gyration fluctuates about 1-5.5 nm, displaying some qualitatively ambiguous periodicity (however this periodicity is likely non-existent). The total number of hydrogen bonds fluctuates stochastically and greatly over 0-25 bonds throughout the 10 ps production run. Both these graphs are measures of the 3D fluctuation of the peptide's structure throughout its production run. Radius of gyration was calculated by comparing the starting structure of the system to the production run structure, with high Rg values indicating large fluctuations away from the initial structure. The hydrogen bond number, as said before, is an indication of the degree of order of the 3D structure. Because the chain of residues that comprise a protein must be bonded together at various points, with hydrogen bonds usually, to organize their 3D conformation, the fact that the number of H-bonds fluctuates so much is clearly indicative of the degree of 3D fluctuation. The magnitude of H-bond flux is also quite low relative to that of an ordered protein. It was seen before that a simple lysozyme simulation fluctuated around 80-100 H-bonds throughout a 20 ns production run, with a clear steady state behaviour around 90 H-bonds. The flux of the H-bond number, as well as the low value of H-bond value throughout production run both indicate the high degree of disorder the eight peptide system possesses throughout the production run.

Now, onto the agreement between the behaviour of the RMSD and time dependent concentration plots. As already established, c(t) follows a tanh behaviour, as seen in (Equation 2). Interestingly enough, this tanh behaviour is mirrored in the RMSD time series. How is it that the root mean square deviation, radius of gyration, hydrogen bond number and thus interconversion of macrostate dynamics perfectly mimic the concentration of folded proteins? Seemingly, deviations in structure all follow the same arctangent behaviour, in that any system can be described as a two-state system. A binary code of "on" or "off"; folded or unfolded; quantum or classical? I wonder, what occurs in that meso-classical space: wave-particle duality? How does a stochastically folding protein, one where the rate of unfolding far outweighs that of folding, mimicthe behaviour of an IDP? Is an IDP not just a protein that unfolds or folds at rates of high discrepant magnitudes? Clearly a two state, stochastically folding protein has identical thermodynamic behaviour to that of a two-state paramagnet: the sigmoid; a universal equation of motion for monotonic functions.

I thus conclude that by altering the folding rate such that it differs by at least two orders of magnitude relative to the unfolding rate, any two state system of a traditional protein can become a disordered protein. What dictates this rate-difference? Is it written in the genetic code, and how? Not chaos, but deterministic stochasticity. I hypothesize that given the time series for volume, pressure, and temperature, each would be a hyperbolic tangent. Comparing the simulated data from the Gillespie algorithm and that of Molecular Dynamics Simulations, it's clear the same dynamics can be achieved with very different computational systems: as binary states, anything, anything, can be decomposed into two states, even intrinsically disordered proteins.

**Conclusion**

In conclusion, a stochastically folding, two state system as simulated by the Gillespie algorithm efficiently emulates the behaviour of an IDP when the rate of unfolding trumps the rate of folding by at least two orders of magnitude. The arctangent behaviour remains, however, the thermodynamic flux as time approaches infinity starts to increase proportionally in terms of magnitude relative to time. In space, the 3D conformation's stochastic fluctuations increase monotonically. Thus, energy is conserved, for simple derivations on Einstein's $E=mc^2$ reveal that as long as the time derivative of and indeed the stochasticity itself are both monotonic functions, the time derivative of energy will be constant and thus an axis of homogeneity. Further work to be done to explore the tie between simulation-based grit and wild type vs carcinogenic: use a biological system with a catalyst to speed up the rate of unfolding in a traditional folded protein: just how close can we get this lysozyme to 8-residue peptide? Just how far can we go with applying the two worlds of Newtonian and non-Newtonian, meso-quantum, mesoscopic, to IDPs? Only time will tell.

**Relevant Equations**

$$\frac{dc}{dt} = -k_{unfold}(c^2) + \frac{1}{V}k_{fold} \quad \text{(Equation 1)}$$

$$c(t) = \sqrt{\frac{1}{v}}\tanh\left(\frac{k_{fold}(t)}{\sqrt{v}}\right) \quad \text{(Equation 2)}$$

$$E = mc^2 \quad \text{(Equation 3)}$$

Xian, R. R. (2020, March 27). PHY478 Final Report: An Introduction to PM2.5s, their Importance, and a Cluster Methodology to Analyze their Meteorological Dynamics .

**Figures**

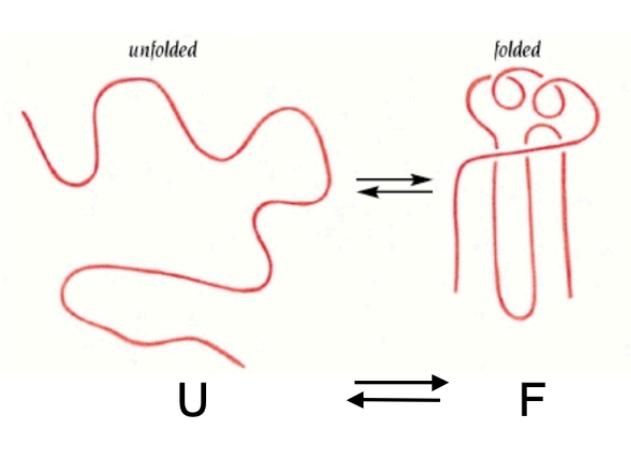

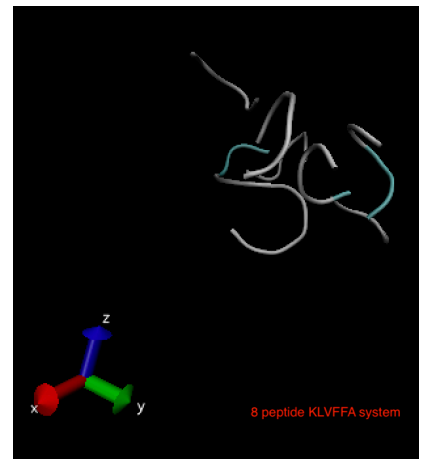

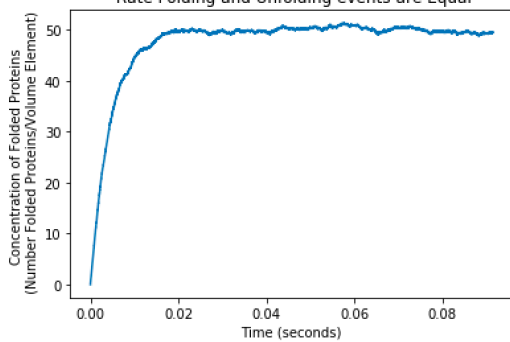

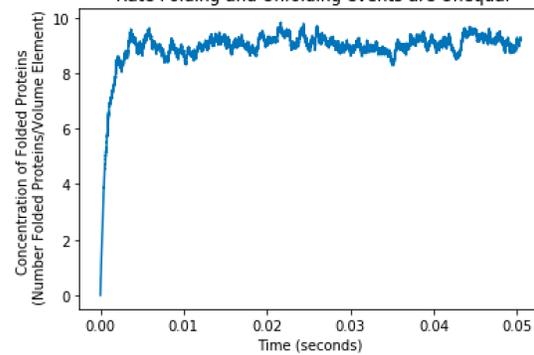

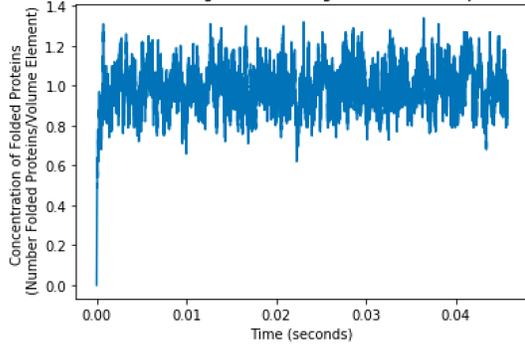
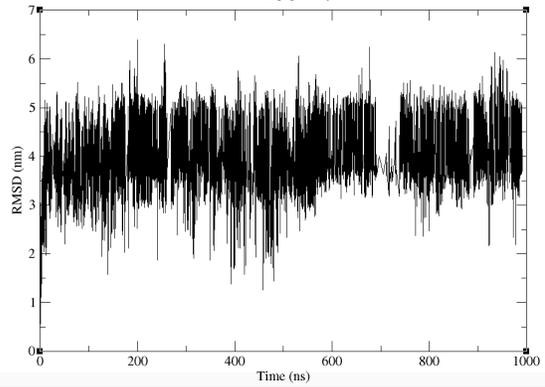
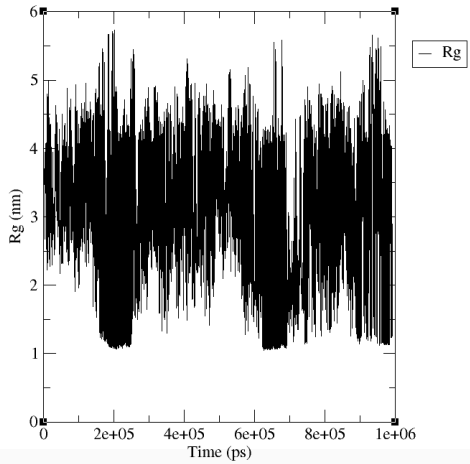
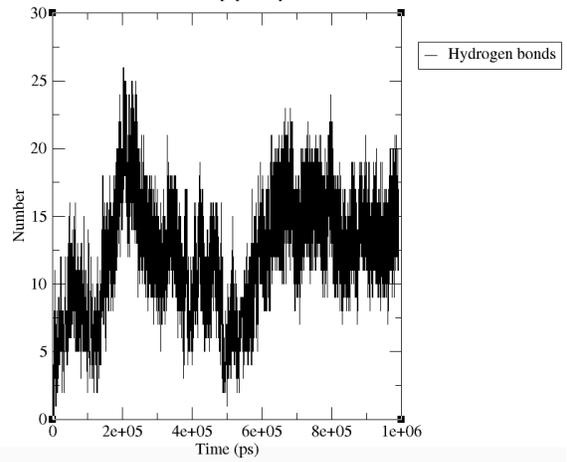
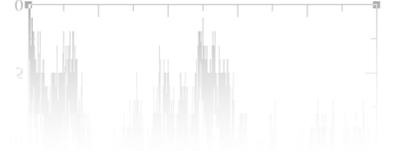